\documentclass{elsart}
\usepackage{graphicx}
\usepackage{amsmath}

\newcommand{\abs}[1]{\left\vert#1\right\vert}

\newcommand{\qt}[1]{``#1''}

\newcommand{\avg}[1]{\left\langle#1\right\rangle}
\newcommand{\keyw}[1]{{\bf #1}}
\renewcommand{\vec}[1]{\ensuremath{\mathbf{#1}}}

\begin{document}
\runauthor{Stramaglia et al.}
\begin{frontmatter}
\title{Statistical mechanics approach to the phase unwrapping problem
}
\author[Dip]{Sebastiano Stramaglia}
\author[INFM,Dip]{Alberto Refice}
\author[Dip]{Luciano Guerriero}

\address[Dip]{Dipartimento Interateneo di Fisica, Via Amendola, 173, Bari (Italy)}
\address[INFM]{Istituto Nazionale di Fisica della Materia (INFM), Sez. G, Bari (Italy)}

\begin{abstract}
The use of Mean-Field theory to unwrap principal phase patterns has 
been recently proposed. In this paper we generalize the Mean-Field 
approach to process phase patterns with arbitrary degree of 
undersampling. The phase unwrapping problem is formulated as that of 
finding the ground state of a locally constrained, finite size, spin-L 
Ising model under a non-uniform magnetic field. The optimization 
problem is solved by the Mean-Field Annealing technique. Synthetic 
experiments show the effectiveness of the proposed algorithm. 
\end{abstract}


\end{frontmatter}


\section{Introduction}
\label{sec:intro}

The determination of the absolute phase from a fringe pattern is an
important problem that finds applications in many areas: homomorphic
signal processing~\cite{Oppenheim-75}, solid state
physics~\cite{Hjalmarson-85}, holographic
interferometry~\cite{Nakadate-85}, adaptive or compensated
optics~\cite{Fried-77}, magnetic resonance imaging~\cite{Ching-92} and
synthetic aperture radar interferometry~\cite{Zebker-86}.

In all these applications one obtains a two-dimensional fringe pattern
whose spatially-varying phase is related to the physical quantity to be
measured. The computation of phase by any inverse trigonometric
function (e.g.\ arctangent) provides only principal phase values, which
lie between $\pm \pi$ radians. The process of phase unwrapping (PU), 
i.e.\ the addition of a proper integer multiple of $2 \pi$ to all the 
pixels, must be carried out before the physical quantity can be 
reconstructed from the phase distributions. Since many possible 
absolute phase fields are compatible with a given fringe pattern, phase 
unwrapping is ill-posed in a mathematical sense: 
Hadamard~\cite{Hadamard-02} defined a mathematical problem to be 
well-posed if a unique solution exists that depends continuously on the 
data; in this case, the uniqueness requirement is violated. 

Ill-posed problems arise frequently in many areas of science and 
engineering; well-known examples are analytic continuation, the Cauchy 
problem for differential equations, computer tomography, and many 
problems in image processing and machine vision that involve the 
reconstruction of images from noisy data. 

The fact that a problem is not well-posed does not mean that it cannot 
be solved: rather, in order to be solved it must be first 
\emph{regularized} by introducing additional constraints (prior 
knowledge) about the behaviour of the solution. Variational 
regularization corresponds to modeling the physical constraints of the 
problem by a suitable functional; the solution is then sought as the 
minimizer of this functional. 

In the last years, an increasing interest has been devoted to adapt 
methods from Statistical Mechanics  to nonconvex optimization problems 
arising from the variational regularization of ill-posed problems. 
Geman and Geman~\cite{Geman-84} suggested that the Ising model is 
applicable to image restoration through the Bayesian formalism. This 
problem corresponds to searching the ground state of a finite-size 
Ising model under a non-uniform external field. Geman and Geman applied 
this to the recovery of corrupted images by using simulated annealing 
of a spin-S Ising model. After that, Gidas~\cite{Gidas-89} proposed a 
new method based on a combination of the renormalization group 
technique and the simulated annealing procedure; then, 
Zhang~\cite{Zhang-92} introduced Mean-Field Annealing to treat the 
image reconstruction problem, while Tanaka and 
Morita~\cite{Tanaka-95,Morita-96} applied the cluster variation method. 
Methods of statistical mechanics have also been used to study 
combinatorial optimization problems (see, e.g.,~\cite{Ray-97} and 
references therein). 

In a couple of recent papers, the phase unwrapping problem was handled 
by methods of Statistical Mechanics. Simulated annealing was applied 
in~\cite{Guerriero-98}. In~\cite{Stramaglia-99} the problem is solved 
by the Mean-Field Annealing (MFA) technique; PU is formulated as a 
constrained optimization problem for the field of integer corrections 
to be added to the wrapped phase gradient in order to recover the true 
phase gradient, with the cost function consisting of second order 
differences, and measuring the smoothness of the reconstructed phase 
field. This is equivalent~\cite{Stramaglia-99} to finding the ground 
state of a locally-constrained ferromagnetic spin-1 Ising model under a 
non-uniform external field. The optimization problem is then solved by 
MFA and consistent solutions are found in difficult situations 
resulting from noise and undersampling. 

Mean Field Annealing is closely related to Simulated 
Annealing~\cite{Geman-84}. Both approaches formulate the optimization 
problem in terms of minimizing a cost function and defining a 
corresponding Gibbs distribution. Simulated Annealing then proceeds by 
sampling the Gibbs probability distribution as the temperature is 
reduced to zero, whereas MFA attempts to track an approximation of the 
mean of the same distribution. 

The algorithm described in~\cite{Stramaglia-99} was constructed under 
the assumption that the possible values for the correction field were 
restricted to belong to the set $\{ -1, 0, 1 \}$. In this paper we 
generalize the MFA algorithm to the case of a set $\{-L, \ldots, L\}$, 
with $L$ an arbitrary integer. The corresponding statistical system is 
a locally-constrained spin-L Ising model. 

The present generalization allows the processing of input phase
patterns with arbitrary degree of undersampling; our experiments on
synthetic phase fields show the effectiveness of the proposed
algorithm.

The paper is organized as follows. In sect.~\ref{sec:pu} the phase 
unwrapping problem is introduced and its ill-posedness is highlighted. 
In sect.~\ref{sec:alg} the deterministic MFA algorithm is described in 
detail. Then, in sect.~\ref{sec:exp} some experimental results on 
simulated phase fields are presented. Some conclusions are then drawn 
in sect.~\ref{sec:concl}.

\section{Phase Unwrapping Problem}
\label{sec:pu}

We briefly recall here the phase unwrapping terminology, and refer the
reader to~\cite{Ghiglia-98} for a complete discussion.

Given an absolute phase pattern $f(x,y)$ on a two-dimensional square 
grid, what is actually measured is the wrapped phase field $g(x,y)$ 
which can be expressed in terms of the $f$ field through a wrapping 
operator, Wr, defined so that $g(x,y)$ always lies in the interval 
$[-\pi,+\pi)$: 
    \begin{equation}
	 g(x,y) = \mathrm{Wr} [f(x,y)] = \arg \left\{ \exp [ \mathrm{i}
	 f(x,y)]\right\}.
    \end{equation}

Phase unwrapping means recovering the absolute phase field $f$, which 
is usually related to the physical quantity to be measured, from the 
knowledge of the $g$ field. This can be done in practice by estimating 
the absolute phase gradient from the wrapped phase field and 
integrating it throughout the 2-D grid. This simple method is effective 
only in absence of phase aliasing, i.e.\ if the phase field is 
correctly sampled. 

In fact, if the Nyquist condition:
    \begin{equation}
	 \abs{\vec{\nabla} f(x,y)} < \pi,
    \label{eq:CondNoAlias}
    \end{equation}
where $\nabla$ is the discrete gradient, is verified everywhere on the 
grid, the absolute phase gradient is obtained by wrapping the gradient 
of the wrapped phase field, according to the formula: 
    \begin{equation}
	 \vec{A}(x,y) = \mathrm{Wr} [ \vec{\nabla} g (x,y)].
    \end{equation}
As mentioned, if condition~\eqref{eq:CondNoAlias} is satisfied, one 
has:
    \begin{equation}
	 \vec{\nabla} f(x,y) = \vec{A}(x,y),
    \label{eq:noalias}
    \end{equation}
and the $f$-pattern is obtained by integrating $\vec{A}$ along any path
connecting all sites on the grid.

The Nyquist condition is often violated because of undersampling of the 
signal from which the principal phase is extracted. This can result 
either from system noise, or from critical values of the slopes of the 
physical surface which is analyzed through interferometry. For example, 
in the case of SAR interferometry, the surface is the portion of Earth 
imaged from the sensor (usually air- or satellite-borne), while noise 
can arise from sensor thermal electronic motion, or from other sources 
of electronic signal disturbances. 

If the Nyquist condition is not satisfied everywhere on the grid, then 
the wrapped gradient $\vec{A}$ of the wrapped phase field is not 
assured to equal the absolute phase gradient. In this case, a more 
general relation must be written, rather than~\eqref{eq:noalias}, i.e.: 
    \begin{equation}
	 \vec{\nabla} f(x,y) = \vec{A}(x,y) + 2 \pi \vec{k}(x,y),
    \label{eq:aliasgrad}
    \end{equation}
where $\vec{k}(x,y)$ is a vector field of integers. In this case, 
solving PU amounts to finding the correct field $\vec{k}$. 

Phase aliasing conditions imply that the integration of field $\vec{A}$ 
depends on the path. The sources of this nonconservative behaviour are 
detectable by calculating the integral of the field $\vec{A}$ over 
every minimum closed path, i.e.\ the 2$\times$2 square having the site 
$(x,y)$ as a corner: 
    \begin{equation}
	 I(x,y) = \frac{1}{2\pi} \bigl[ A_x(x,y) + A(y(x+1,y) - A_x(x,y+1) - A_y(x,y) 
	 \bigr].
    \label{eq:resdef}
    \end{equation}
One can show that the integral $I(x,y)$ will always have a value in the 
set $\{-1, 0, 1\}$. Locations with $I \neq 0$ are called \qt{residues}. 
In presence of residues, the field $\vec{A}$ is no more irrotational; 
this causes the path-dependence of the integration step previously 
described.

To restore the consistency of the phase gradient, then, the $\vec{k}$ 
field must satisfy the following consistency condition: 
    \begin{equation}
	 \vec{\nabla} \times \left[ \vec{A}(x,y) + 2\pi
	 \vec{k}(x,y) \right] = 0,
    \label{eq:consist}
    \end{equation}
where $\nabla \times \cdot$ is the discrete curl operator. Since there 
are many possible $\vec{k}$ fields satisfying eq.~\eqref{eq:consist}, 
PU is an ill-posed problem according to Hadamard's definition. 

One of the most classical and widely-used algorithms for phase 
unwrapping is the Least Mean Squares (LMS) approach~\cite{Ghiglia-94}, 
which consists in finding the scalar field $f$ whose gradient is closer 
to $\vec{A}$ in the Least Squares sense, i.e.\ the minimizer of: 
	\begin{equation}
		\sum ( \vec{\nabla} f - \vec{A})^2.
	\end{equation}

As mentioned before, in~\cite{Stramaglia-99} a variational approach has
been used, and the field $\vec{k}$ was \qt{chosen} as the minimizer of
the following functional:
    \begin{eqnarray}
    R &=& {1\over 4\pi^2}      \sum 
		    \left[
		    \nabla_{x} f(x+1,y) - \nabla_{x} f(x,y)
		    \right]^2 \nonumber \\ &\qquad& +
    {1\over 4\pi^2}      \sum 
		    \left[
		    \nabla_{y} f(x,y+1) - \nabla_{y} f(x,y) 
		    \right]^2 \nonumber \\
	    &\qquad& + {1\over 4\pi^2}\sum
		    \left[
		    \nabla_{x} f(x,y+1) - \nabla_{x} f(x,y)
		    \right]^2 \nonumber \\ &\qquad& +
	    {1\over 4\pi^2}\sum
		    \left[
		    \nabla_{y} f(x+1,y) - \nabla_{y} f(x,y)
		    \right]^2,
    \label{eq:nostro}
    \end{eqnarray}
subject to constraint~\eqref{eq:consist}. Due to~\eqref{eq:aliasgrad}, 
$R$ is a functional of $\vec{k}$, i.e.\ $R=R[\vec{k}]$, and it measures 
the smoothness of the reconstructed phase surface. The optimization 
problem was then solved by a MFA algorithm under the assumption that 
the $\vec{k}$ field be restricted to take values in $\{-1, 0, 1\}$. In 
the next section we generalize the MFA algorithm to the case of 
$\vec{k}$ fields belonging to $\{-L, \ldots, L\}$, with $L$ an 
arbitrary integer. 

\section{The algorithm}
\label{sec:alg}

As explained in sect.~\ref{sec:pu}, we assume the solution of PU to be 
the minimizer of the functional~\eqref{eq:nostro}, subject to 
constraint~\eqref{eq:consist}. Let us assume that the possible values 
of the $\vec{k}$ field are restricted to belong to $\{-L, \ldots, L\}$. 
The field $\vec{k}$ may then be regarded as a system of spin-L units. 
We assume the Gibbs distribution: 
	\begin{equation}
		P[\vec{k}] = \frac{\exp\left[ \frac{-R[\vec{k}]}{T} \right]}
		{\sum_{\vec{k}'} \exp \left[ \frac{-R[\vec{k}']}{T} 
		\right]},
	\end{equation}
where the sum is over the $\vec{k}'$ fields 
satisfying~\eqref{eq:consist}, and $T$ is the statistical temperature. 
Inconsistent fields $\vec{k}'$ are assumed to have zero probability. 

Following Mean-Field theory~\cite{Parisi-88}, we consider a probability 
distribution for the correction field $\vec{k}$ which treats all the 
variables as independent, i.e.\ it is the product of the marginal 
distributions of each variable. 

Let $\rho_x(x,y,\alpha)$ be the probability that $k_x(x,y) = \alpha$,
with $\alpha = -L, \ldots, L$, and $\rho_y(x,y,\alpha)$ be the
corresponding probability for $k_y(x,y)$. Normalization of these
marginal probabilities implies a penalty functional:
    \begin{equation}
	 \Theta[\rho] = \sum_{(x,y)} \biggl[ V_x(x,y) \left(1 -
	 \sum_{\alpha} \rho_x(x,y,\alpha) \right) +
	 V_y(x,y) \left(1 -
	 \sum_{\alpha} \rho_y(x,y,\alpha) \right) \biggr],
    \end{equation}
where $\{V\}$ are Lagrange multipliers.

The entropy of the system, in the mean field approximation, is:
    \begin{equation}
	 S[\rho] = - \sum_{(x,y)} \sum_{\alpha = -L}^L \bigl[
	 \rho_x(x,y,\alpha) \log \rho_x(x,y,\alpha) + \rho_y(x,y,\alpha)
	 \log \rho_y(x,y,\alpha) \bigr].
    \end{equation}

It is useful to introduce the average fields $\vec{m} =
\avg{\vec{k}}_{\vec{\rho}}$ and $\vec{Q} =
\avg{\vec{k}^2}_{\vec{\rho}}$, defined by:
    \begin{eqnarray}
	 m_x(x,y) = \sum_{\alpha = -L}^L \alpha \rho_x (x,y,\alpha),
	 \qquad m_y(x,y) = \sum_{\alpha = -L}^L \alpha \rho_y
	 (x,y,\alpha);
	 \label{eq:avefields1}\\
	 Q_x(x,y) = \sum_{\alpha = -L}^L \alpha^2 \rho_x (x,y,\alpha),
	 \qquad Q_y(x,y) = \sum_{\alpha = -L}^L \alpha^2 \rho_y
	 (x,y,\alpha).
	 \label{eq:avefields2}
    \end{eqnarray}
The average $U$ of the cost functional $R$ is called \emph{internal
energy}. It is easy to show that the internal energy depends only on
$\vec{m}$ and $\vec{Q}$:
    \begin{equation}
	 U[\vec{m},\vec{Q}] = \avg{R[\vec{A} + 2 \pi
	 \vec{k}]}_{\vec{\rho}}.
    \end{equation}

A penalty functional is introduced to enforce 
constraints~\eqref{eq:consist}:
	\begin{eqnarray}
		\Gamma [ \vec{m} ] &= \sum_{(x,y)} \lambda (x,y) \bigl[ 
		m_x(x,y) + m_y(x+1,y) \nonumber \\
		&\qquad - m_x(x,y+1) - m_y(x,y) + I(x,y) 
		\bigr],
	\end{eqnarray}
where $\{\lambda\}$ is another set of Lagrange multipliers.

Let us now introduce an effective cost functional, the \emph{free 
energy}, which depends on $T$:
	\begin{equation}
		F[\rho] = U[\vec{m}, \vec{Q}] - T S[\vec{\rho}] + \Gamma[\vec{m}] + 
		\Theta[\vec{\rho}]
	\end{equation}

The free energy is the weighted sum of the internal energy (the 
original cost function) and the entropy functional; $\Gamma$ and 
$\Theta$ are penalty functionals to enforce the constraints of the 
problem. According to the variational principle of Statistical 
Mechanics, the best approximation to the Gibbs distribution is the 
minimizer of the free energy~\cite{Parisi-88}. Since $-TS$ is a convex 
functional, the free energy is convex at high temperature and the 
global minimum can be easily attained. The solution can then be 
continuated as temperature is lowered, so as to reach a minimum of $U$. 
This procedure has shown to be less sensitive to local minima than 
conventional descent methods, and gives results close to the ones from 
Simulated Annealing, while requiring less computational 
time~\cite{Yuille-94}. 

The equations for the minimum of the free energy are usually called 
\qt{mean-field equations}:
	\begin{equation}
		\frac{\partial F}{\partial \rho_x (x,y,\alpha)} = 0; 
		\quad \frac{\partial F}{\partial \rho_y (x,y,\alpha)} = 0
	\label{eq:mf}
	\end{equation}

After simple calculations, the solution of eqs.~\eqref{eq:mf} is found 
to have the following form:
	\begin{eqnarray}
		\rho_x(x,y,\alpha) &= 
			\frac{\exp\left\{ -\beta \left[ 
				\frac{\partial U}{\partial \rho_x(x,y,\alpha)} +  
				\frac{\partial \Gamma}{\partial \rho_x(x,y,\alpha)}
				\right] \right\}}
			 {\sum_{\alpha' = -L}^L \exp \left\{ -\beta \left[ 
				\frac{\partial U}{\partial \rho_x(x,y,\alpha')} +  
				\frac{\partial \Gamma}{\partial \rho_x(x,y,\alpha')}
				\right] \right\}},
		\label{eq:solutionrho1}\\
		\rho_y(x,y,\alpha) &= 
			\frac{\exp\left\{ -\beta \left[ 
				\frac{\partial U}{\partial \rho_y(x,y,\alpha)} +  
				\frac{\partial \Gamma}{\partial \rho_y(x,y,\alpha)}
				\right] \right\}}
			 {\sum_{\alpha' = -L}^L \exp \left\{ -\beta \left[ 
				\frac{\partial U}{\partial \rho_y(x,y,\alpha')} +  
				\frac{\partial \Gamma}{\partial \rho_y(x,y,\alpha')}
				\right] \right\}}
		\label{eq:solutionrho2}
, 
	\end{eqnarray}
where the $\{V\}$ multipliers have been fixed to normalize the 
distributions, and $\beta = 1/T$ is the inverse temperature. Now we 
observe that, for each site $(x,y)$ on the grid:
	\begin{equation}
		\frac{\partial U}{\partial \rho_x(\alpha)} = 
			\frac{\partial U}{\partial m_x} \frac{\partial m_x}{\partial \rho_x(\alpha)}
			+ \frac{\partial U}{\partial Q_x} \frac{\partial Q_x}{\partial 
			\rho_x(\alpha)} =
			\alpha \frac{\partial U}{\partial m_x(x,y)} + 
			\alpha^2 \frac{\partial U}{\partial Q_x(x,y)}.
	\label{eq:partial1}
	\end{equation}

Analogously, one can easily find:
	\begin{eqnarray}
	\label{eq:partial2}
		\frac{\partial U}{\partial \rho_y(\alpha)} &=& 
			\alpha \frac{\partial U}{\partial m_y(x,y)} + 
			\alpha^2 \frac{\partial U}{\partial Q_y(x,y)},\\
		\frac{\partial \Gamma}{\partial \rho_x(\alpha)} &=&
			\alpha \left[ \lambda(x,y) - \lambda(x,y-1) 
			\right],\\
		\frac{\partial \Gamma}{\partial \rho_y(\alpha)} &=& 
			\alpha \left[ - \lambda(x,y) + \lambda(x-1,y) 
			\right].
	\end{eqnarray}

The derivatives of $U$ with respect to the $\{m\}$ and $\{Q\}$ 
variables are reported in Appendix A. From these expressions it is 
clear that the present formulation of PU is equivalent to finding the 
ground state of a finite-size, spin-L Ising model with local 
constraints, and under a non-uniform magnetic field. 

The consistency constraints are written as equations for the 
$\{\lambda\}$ field:
	\begin{eqnarray}
		\lambda(x,y) &= \lambda(x,y) - 
			b \bigl[ m_x(x,y) + m_y(x+1,y) \nonumber \\
			&\qquad - m_x(x,y+1) - m_y(x,y) + I(x,y) 
			\bigr],
	\label{eq:solutionlambda}
	\end{eqnarray}
where $b$ is a small constant.

Equations~(\ref{eq:solutionrho1}--\ref{eq:solutionrho2}) 
and~\eqref{eq:solutionlambda} are the mean-field equations for PU for 
arbitrary $L$. As already explained, the MFA technique consists in 
solving iteratively the mean-field equations at high temperature (low 
$\beta$), and then track the solution as the temperature is lowered 
($\beta$ grows). 

The algorithm can be summarized as follows. The initial distributions 
give the same probability to each value of the correction field, i.e.\ 
$\rho_x(x,y,\alpha) = \rho_y(x,y,\alpha) = \frac{1}{2L+1}$; the inverse 
temperature is set to $\beta_{\mathrm{MIN}}$ ($\beta_{\mathrm{MAX}}$ is 
the lowest temperature). Then:
	\begin{enumerate}
		\item \keyw{Evaluate} $\{\vec{m}\}$ and $\{\vec{Q}\}$ 
		fields by 
		Eqs.~(\ref{eq:avefields1}--\ref{eq:avefields2});
		\item \keyw{Iterate} 
		Eqs.~(\ref{eq:solutionrho1}--\ref{eq:solutionrho2});
		\item \keyw{Iterate} 
		Eq.~(\ref{eq:solutionlambda});
		\item \keyw{If} 
		Eqs.~(\ref{eq:solutionrho1}--\ref{eq:solutionrho2}) or 
		Eq.~(\ref{eq:solutionlambda}) are not satisfied, 
		\keyw{goto} step 1;
		\item \keyw{If} $\beta < \beta_{\mathrm{MAX}}$, increase 
		$\beta$ and \keyw{goto} step 1.
	\end{enumerate}

The output of this algorithm is a field $\{\vec{m}_{\mathrm{OUT}}\}$ 
which approximates the average of $\{\vec{k}\}$ over the global minima 
of the cost functional $R$.

We remark that the output of the algorithm described 
in~\cite{Stramaglia-99} satisfies $m \in [-1, 1]$ for each component of 
$\{\vec{m}_{\mathrm{OUT}}\}$, whereas the present algorithm satisfies 
the weaker constraint $m \in [-L, L]$ and therefore can be used also in 
the case of high degree of undersampling. The estimate for the true 
phase gradient is $(\vec{\nabla} f)_{\mathrm{est}} = \vec{A} + 2\pi 
\vec{m}_{\mathrm{OUT}}$; the phase pattern $f$ can then be 
reconstructed by $(\vec{\nabla} f)_{\mathrm{est}}$ as described 
in~\cite{Stramaglia-99}.

\section{Experiments}
\label{sec:exp}

In this section we describe some experiments we performed to test the 
effectiveness of the proposed algorithm.

\begin{figure}
	\begin{center}
		\includegraphics[width=0.6\textwidth]{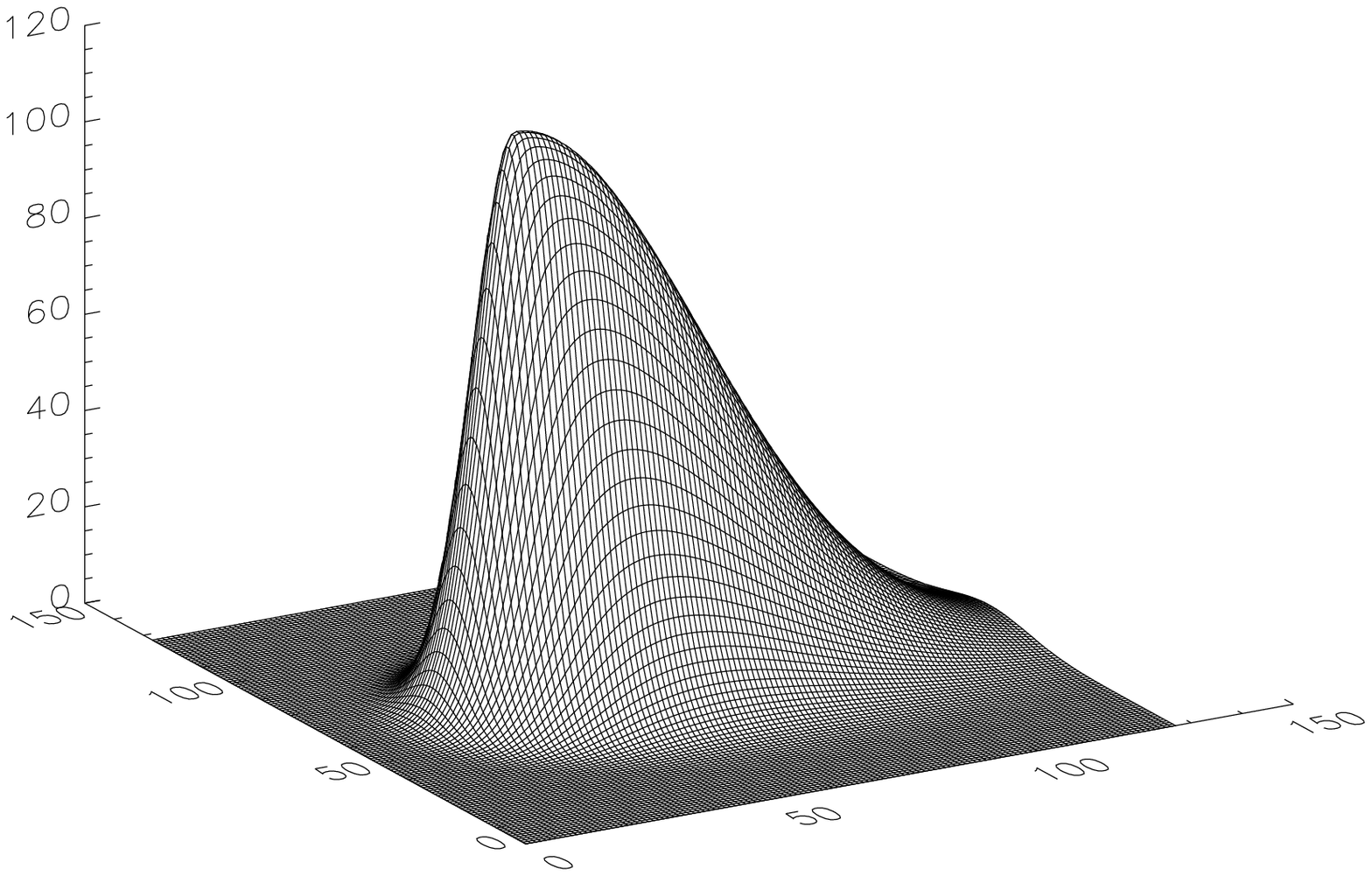}
		\includegraphics[width=0.35\textwidth]{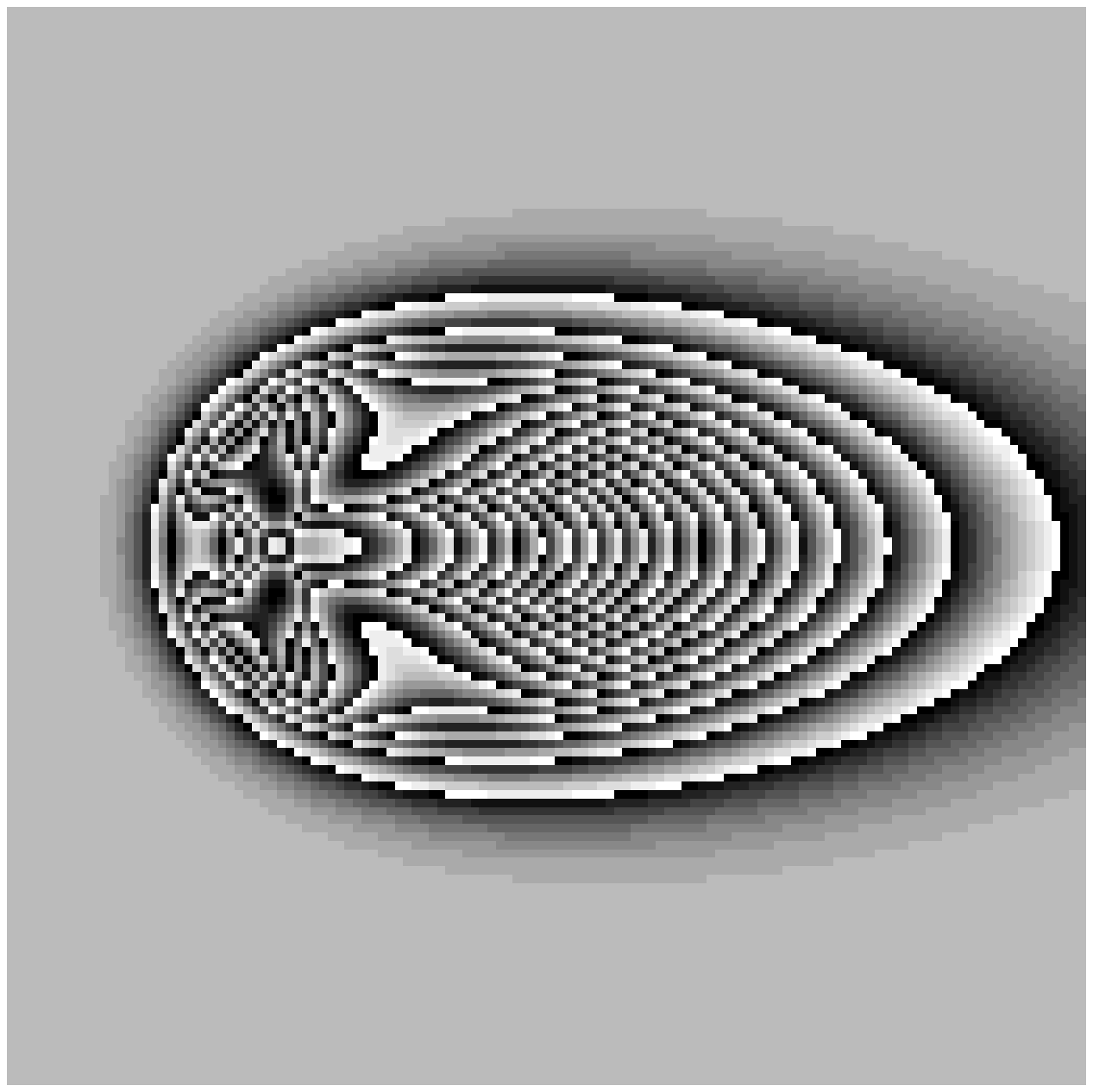}
		\\
		(a) \hspace{0.4\textwidth} (b)
	\end{center}
	\caption{(a) Synthetic phase surface generated via eq.~\eqref{eq:experim}; 
		(b) wrapped phase pattern:
		principal phase values span the interval from $-\pi$ (black pixels) 
		to $+ \pi$ (white pixels).}
	\label{fig:one}
	\end{figure}

In fig.~\ref{fig:one}-(a) a synthetic phase pattern is shown, while in 
fig.~\ref{fig:one}-(b) the wrapped phase pattern is depicted. This test 
phase pattern has been constructed by the following formula: 
	\begin{equation}
		\Phi(x,y) = 120 \exp \left[-\half r^2(x,y) (\mu_1 + \mu_2 c(x,y)) 
		\right], \quad 1 \leq x,y \leq 128
	\label{eq:experim}
	\end{equation}
with:
	\begin{eqnarray}
		r(x,y) &=& \sqrt{(x - 35.5)^2 + (y - 65.5)^2}, \nonumber\\
		c(x,y) &=& \frac{x - 35.5}{r}, \nonumber\\
		\mu_1  &=& 0.01, \nonumber\\
		\mu_2  &=& 0.0004, \nonumber
	\end{eqnarray}
where $\Phi$ is in radians. 

\begin{figure}
	\begin{center}
		\includegraphics[width=0.35\textwidth]{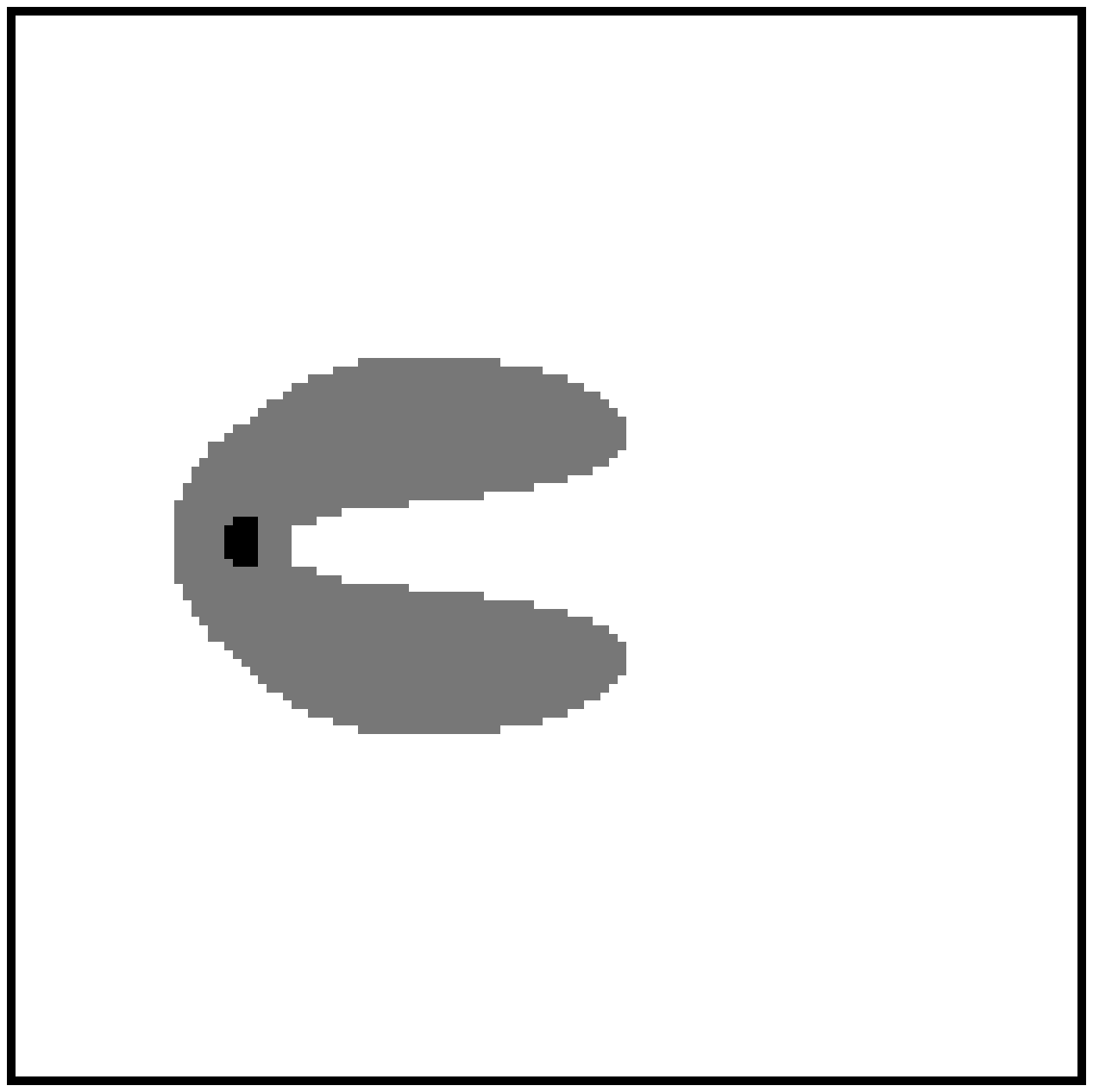}
		\includegraphics[width=0.35\textwidth]{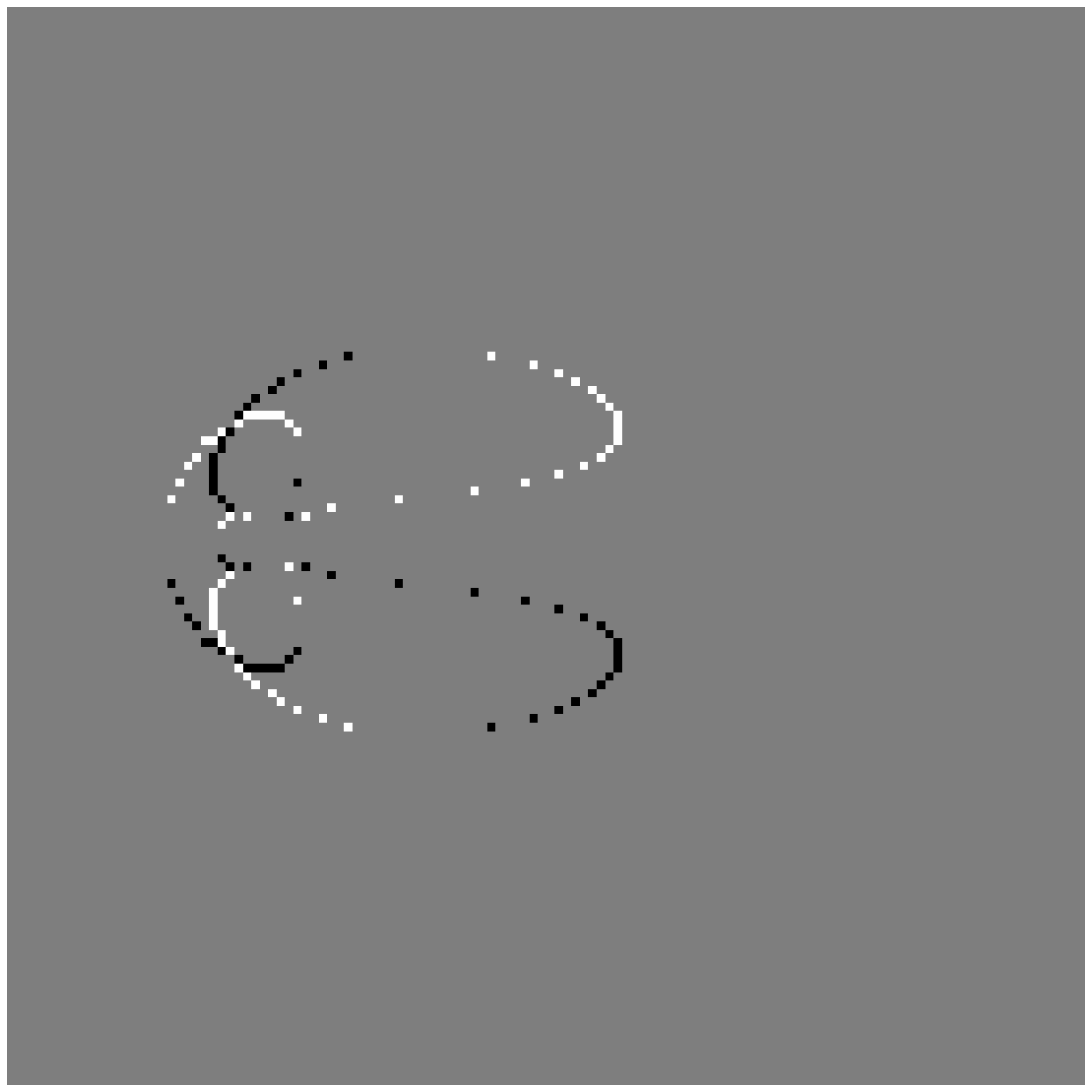}
		\\
		(a) \hspace{0.4\textwidth} (b)
	\end{center}
	\caption{(a) Degree of undersampling of the phase field depicted in 
		fig.~\ref{fig:one}:
		gray pixels represent locations where the absolute phase gradient differs from
		its estimate (wrapping of the principal phase gradient) by one 
		$2\pi$-cycle, black pixels represent locations where the difference is 2 cycles;
		(b) residue map: white pixels are positive residues ($I=1$), 
		black pixels are negative residues ($I=-1$), 
		gray pixels correspond to irrotational locations ($I=0$).}
	\label{fig:two}
	\end{figure}

By construction, $\Phi$ is undersampled: in fig.~\ref{fig:two}-(a) 
black pixels represent locations where the true correction field 
$\vec{k}$ is such that \makebox{$\max \{ k_x, k_y \} = 2$}, while gray 
pixels represent locations where $\max \{ k_x, k_y \} = 1$. The residue 
map is depicted in fig.~\ref{fig:two}-(b). 

\begin{figure}
	\begin{center}
		\includegraphics[width=0.6\textwidth]{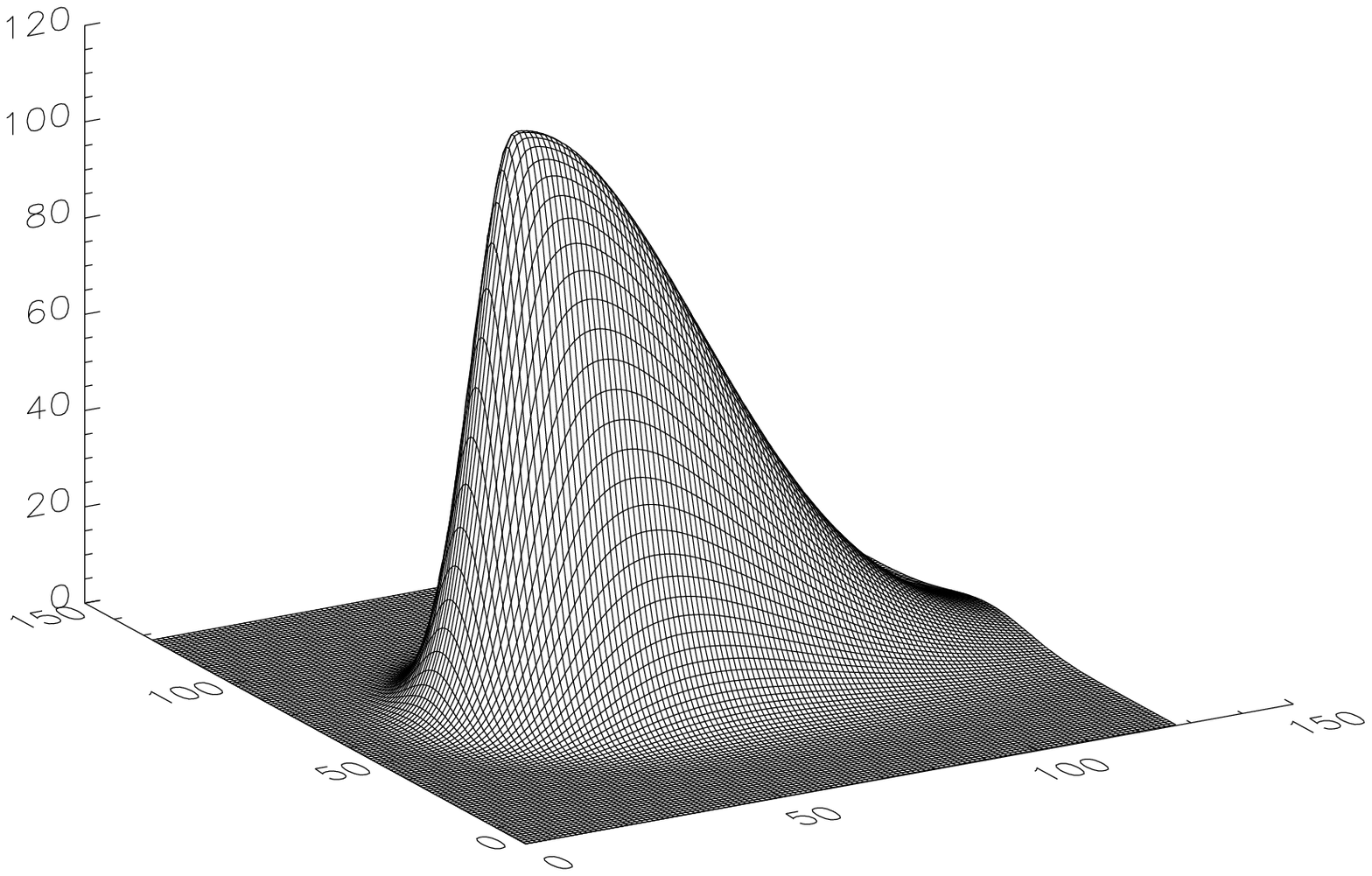}
	\end{center}
	\caption{Unwrapped phase reconstructed by the proposed algorithm 
		from the wrapped phase field shown in fig.~\ref{fig:one}-(b).}
	\label{fig:three}
	\end{figure}

\begin{figure}
	\begin{center}
		\includegraphics[width=0.6\textwidth]{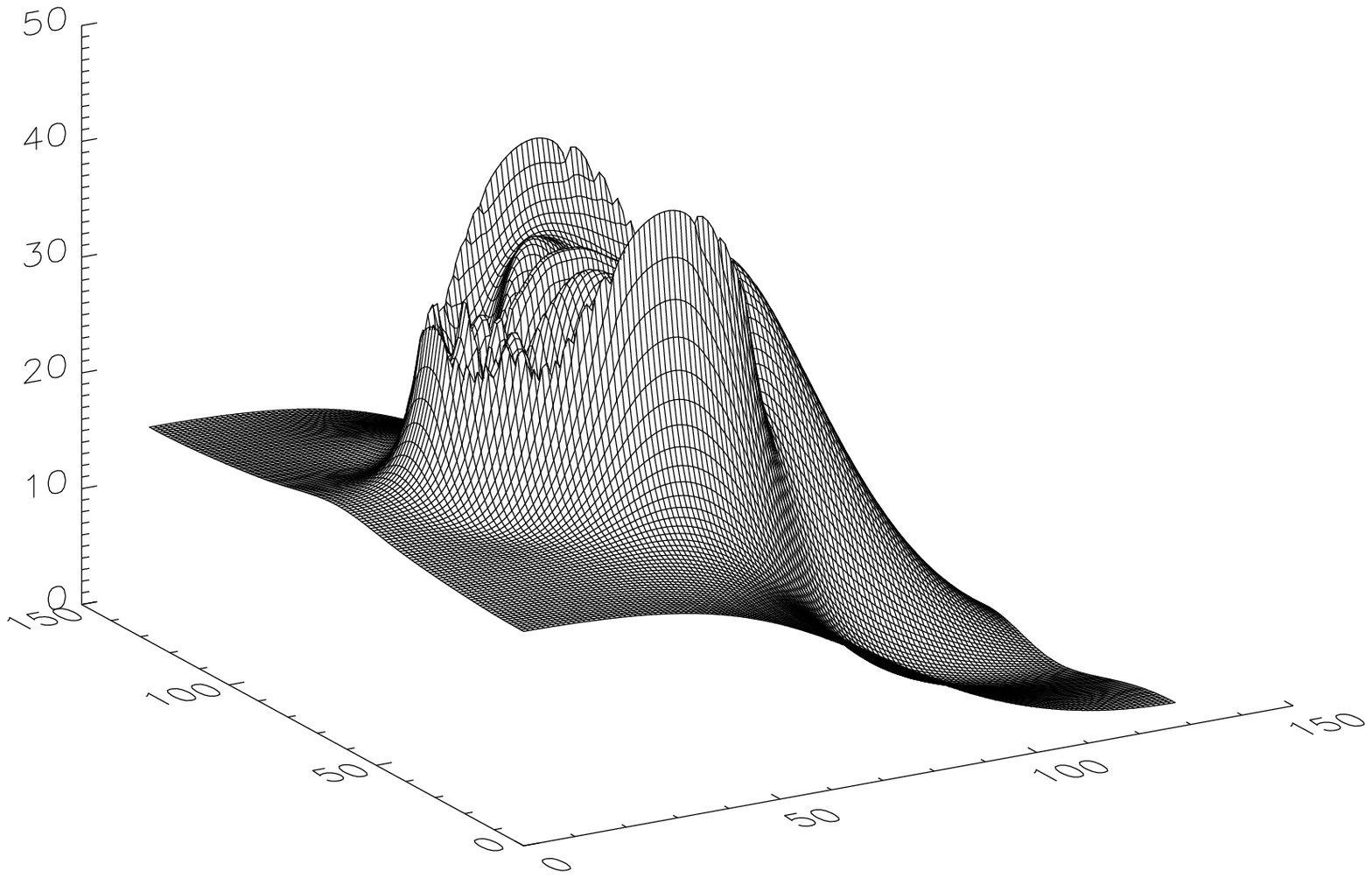}
	\end{center}
	\caption{Unwrapped phase reconstructed by the LMS algorithm 
		from the wrapped phase field shown in fig.~\ref{fig:one}-(b).}
	\label{fig:threethree}
	\end{figure}

We applied the proposed algorithm to unwrap this test pattern. We used 
$L=2, b=0.05$ and the annealing schedule was established as consisting 
of 25 temperature values, equally spaced in the interval 
$[\beta_{\mathrm{MIN}}=0.05, \beta_{\mathrm{MAX}}=1.5]$. The output 
phase pattern is depicted in fig.~\ref{fig:three}. The computational 
time was comparable to that corresponding to~\cite{Stramaglia-99}. The 
input phase surface was perfectly reconstructed. Let us now compare the 
performance of the MFA algorithm with that from LMS~\cite{Ghiglia-94}. 
In fig.~\ref{fig:threethree} we show the output of LMS applied to the 
surface of fig.~\ref{fig:one}-(a). Due to severe undersampling, the LMS 
performance is poor.  

\begin{figure}
	\begin{center}
		\includegraphics[width=0.6\textwidth]{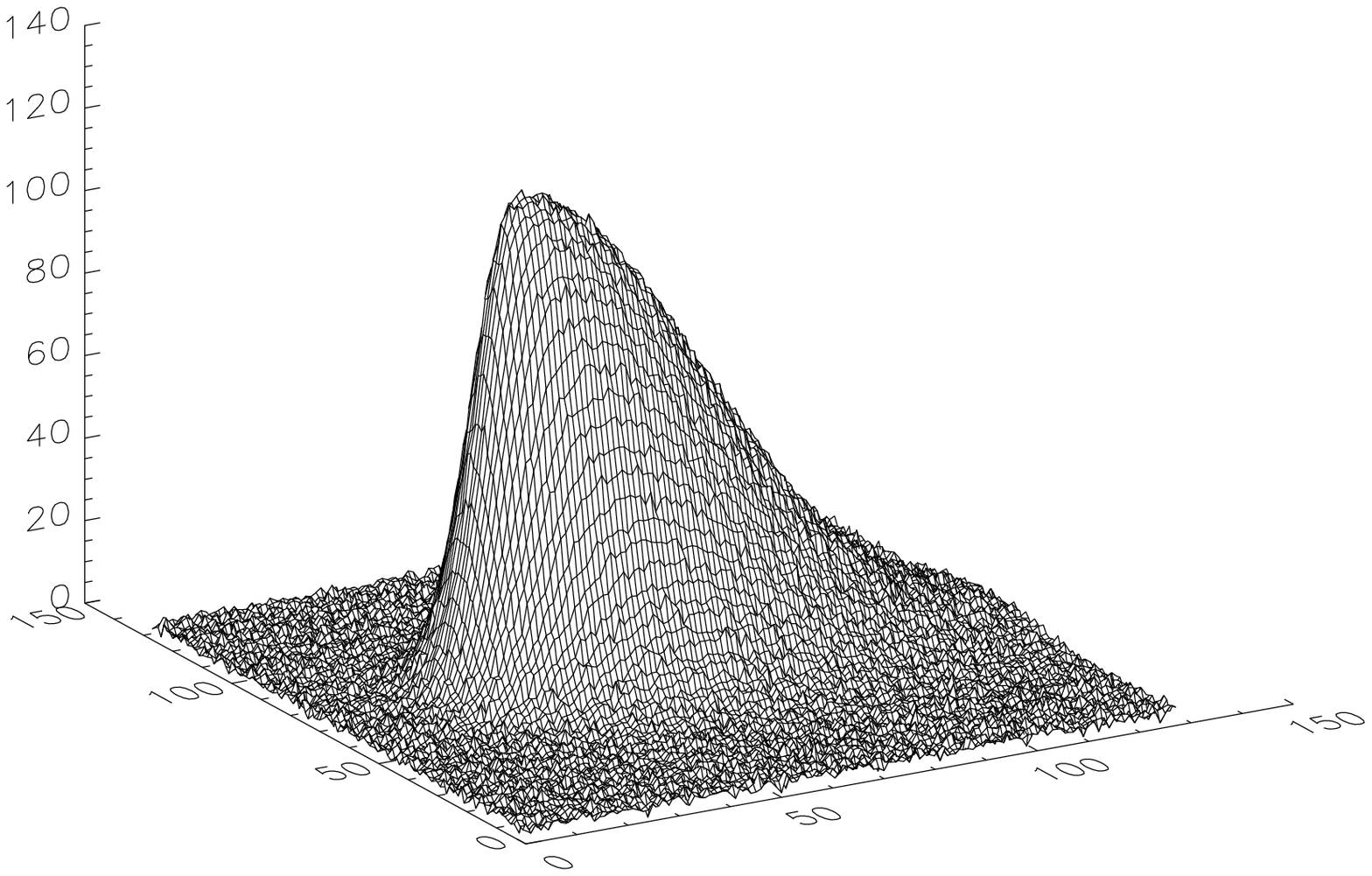}
		\includegraphics[width=0.35\textwidth]{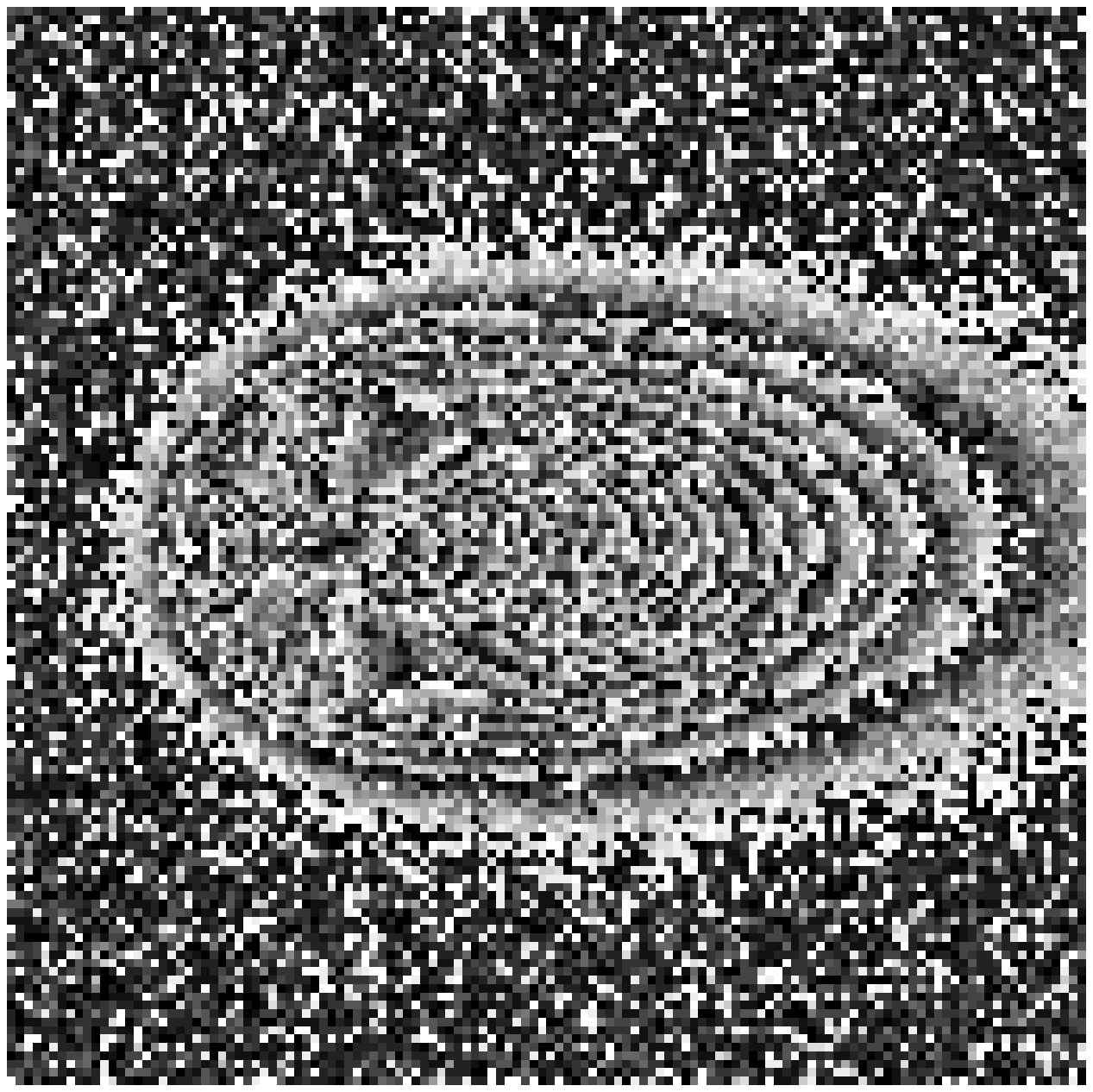}
		\\
		(a) \hspace{0.4\textwidth} (b)\\
		\includegraphics[width=0.35\textwidth]{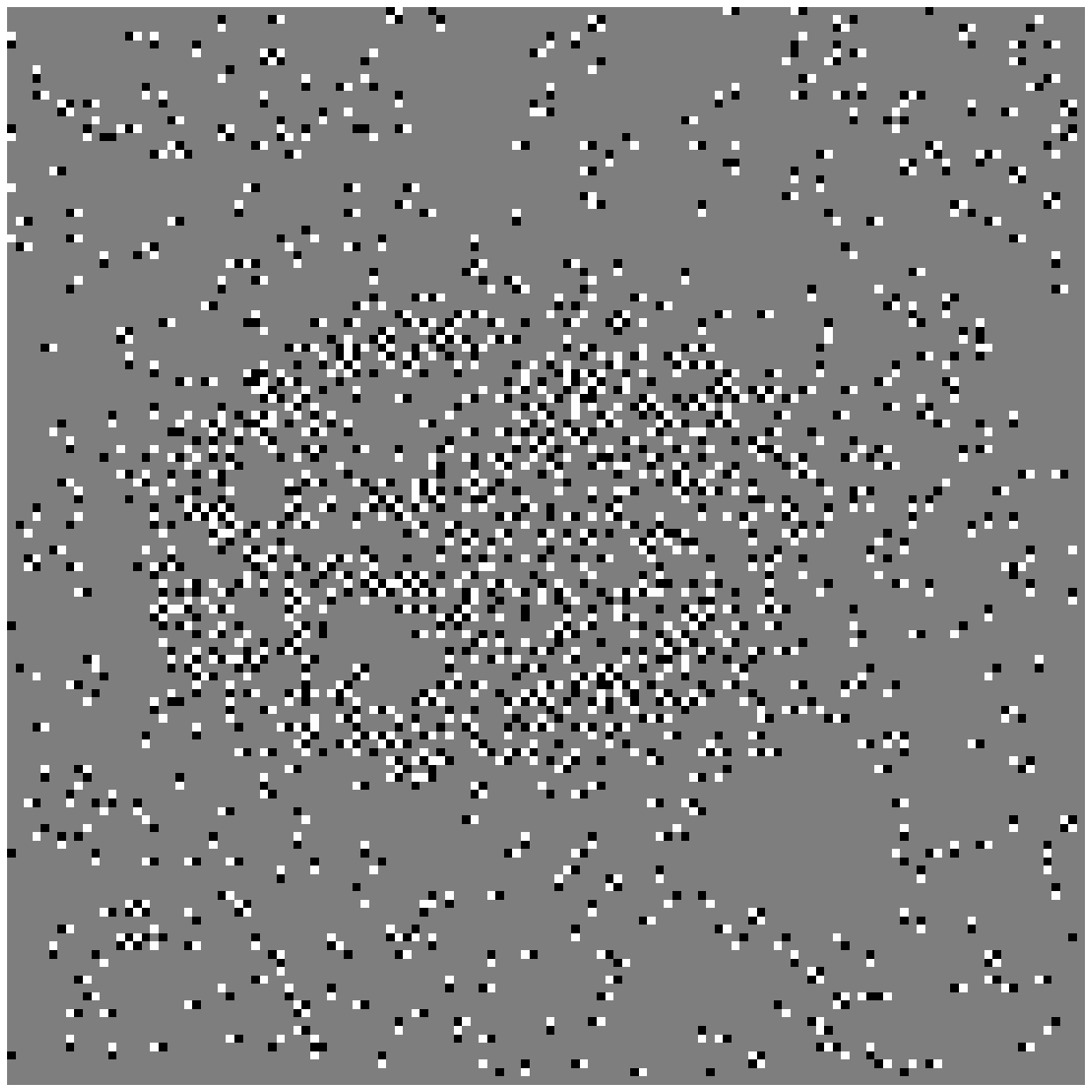}\\
		(c)
	\end{center}
	\caption{(a) Same synthetic surface as in fig.~\ref{fig:one}, 
		with added Gaussian noise with unit variance; (b) principal phase pattern;
		(c) residue map.}
	\label{fig:four}
	\end{figure}

\begin{figure}
	\begin{center}
		\includegraphics[width=0.6\textwidth]{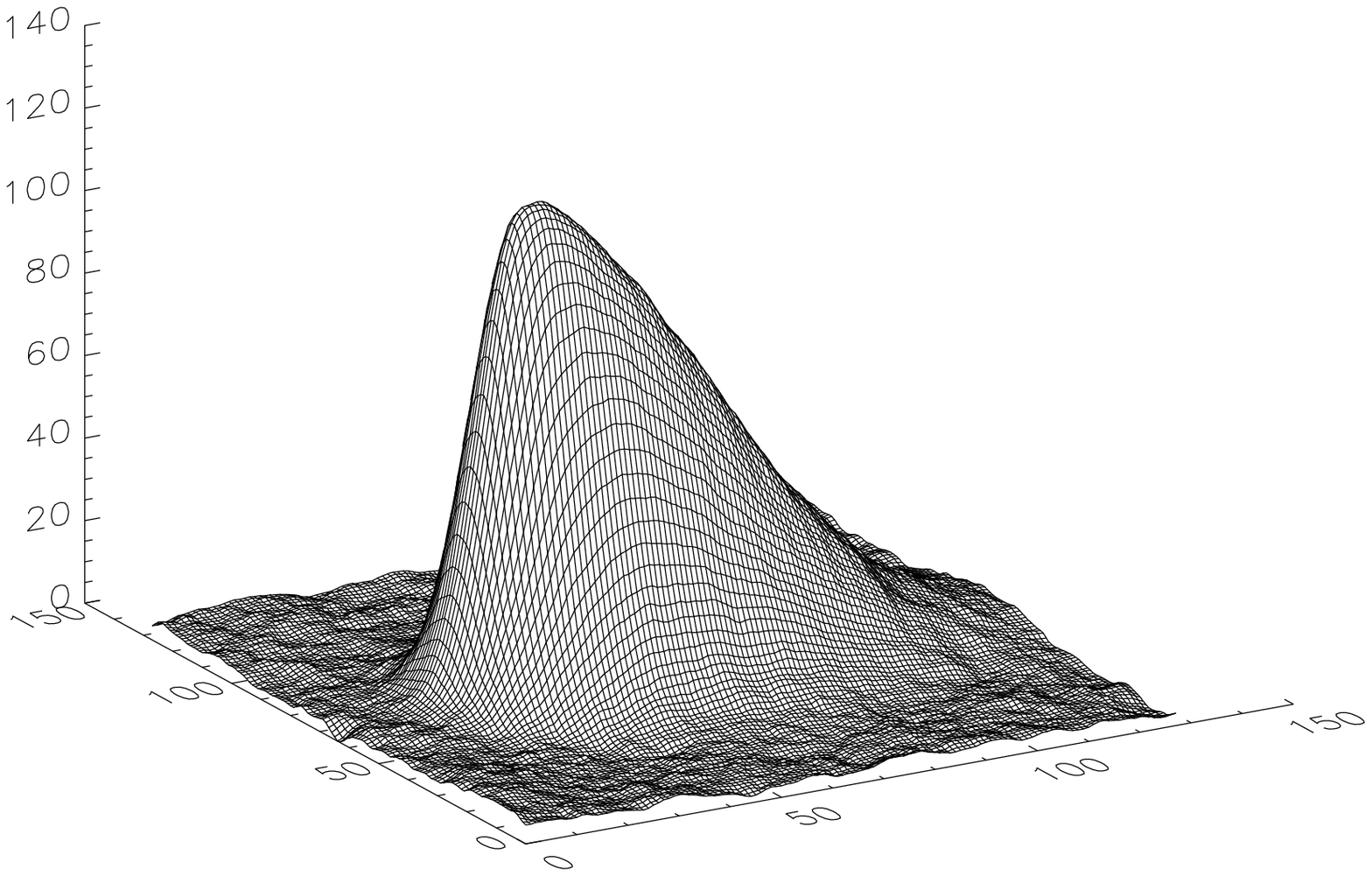}
		\includegraphics[width=0.35\textwidth]{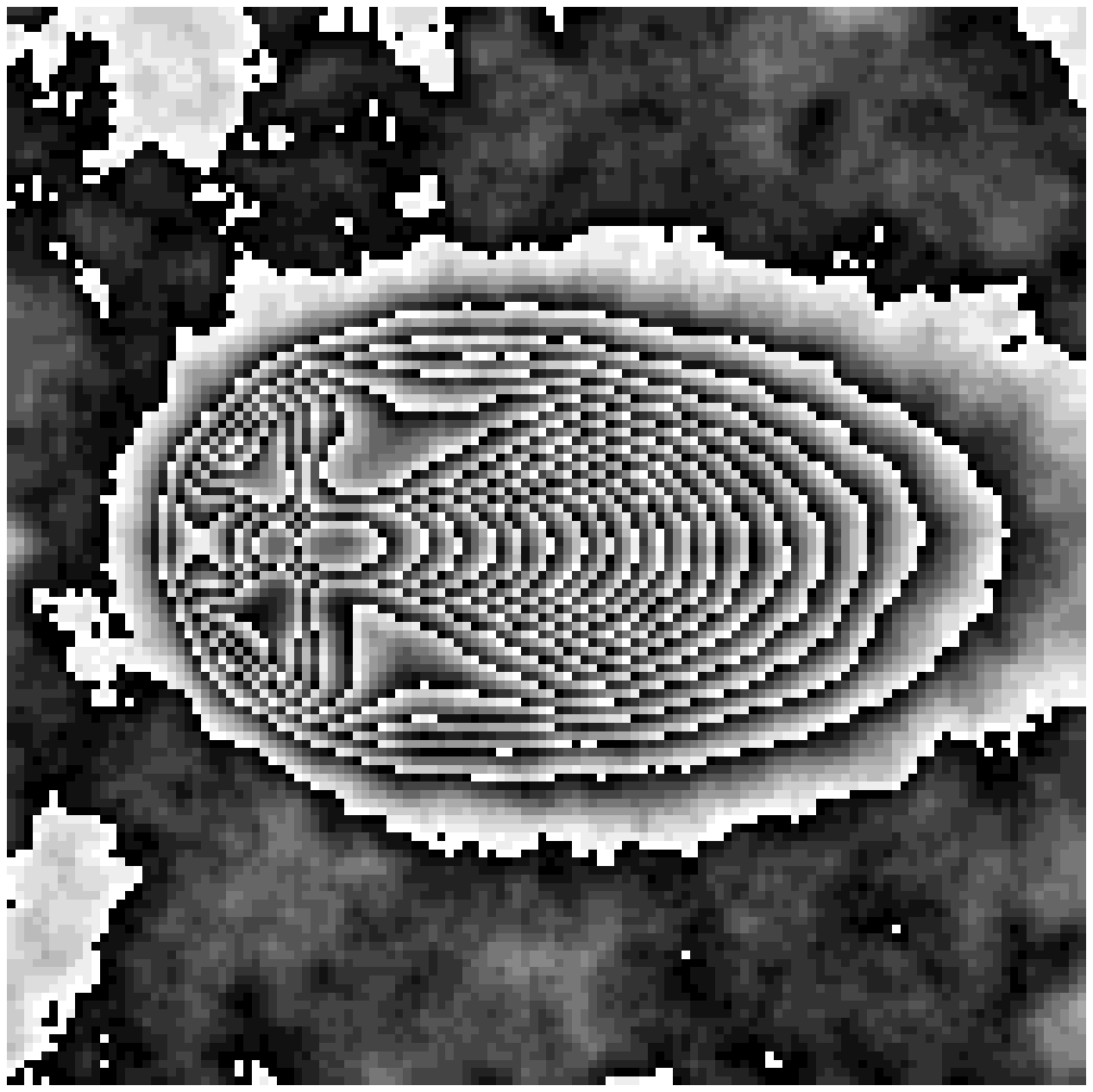}
		\\
		(a) \hspace{0.4\textwidth} (b)
	\end{center}
	\caption{(a) Unwrapped surface reconstructed by the proposed algorithm 
		from the wrapped phase field shown in fig.~\ref{fig:four}-(b);
		(b) principal phase pattern obtained by re-wrapping the solution depicted in (a).}
	\label{fig:five}
	\end{figure}

We also investigated the robustness of the MFA algorithm with respect 
to noise. In fig.~\ref{fig:four}-(a) the phase pattern obtained by 
adding unit-variance Gaussian noise to the surface of 
fig.~\ref{fig:one}-(a) is shown. The wrapped phase pattern is depicted 
in fig.~\ref{fig:four}-(b), while in fig.~\ref{fig:four}-(c) the 
inconsistencies are shown. The output of the MFA algorithm is shown in 
fig.~\ref{fig:five}-(a), while in fig.~\ref{fig:five}-(b) we show the 
phase pattern obtained by re-wrapping the MFA output. The smoothing 
capability of the proposed algorithm appears clearly by comparing 
figs.~\ref{fig:four}-(b) and~\ref{fig:five}-(b). 

\section{Conclusions}
\label{sec:concl}

In this paper we have generalized a previously presented MFA algorithm 
to unwrap phase patterns. This problem is formulated as that of finding 
the ground state of a locally-constrained, spin-L Ising model under a 
non-uniform external field. The present generalization allows 
processing of noisy and highly undersampled input phase fields. The 
effectiveness of this statistical approach to PU has been demonstrated 
on simulated phase surfaces. Further work will be devoted to the 
estimation of the optimal value of $L$ from the observed wrapped phase 
data.

\clearpage
	%
	%

\clearpage
\appendix
\section{Appendix: Derivatives of the internal energy}
\label{sec:app} 

We report here the expressions of the derivatives of the internal 
energy in the Mean-field approximation. 

One easily finds that:
\begin{eqnarray}
  {\partial U\over \partial Q_x(x,y)}=4 \nonumber \\ 
  {\partial U\over \partial Q_y(x,y)}=4, 
				\label{s1}
\end{eqnarray}

\begin{eqnarray}
   {\partial U\over \partial m_x(x,y)}&=& -2 m_x(x-1,y)+{1\over 
   \pi}\left[A_x(x,y)-A_x(x-1,y)\right] + \nonumber \\ &&  -2 
   m_x(x+1,y)+{1\over \pi}\left[A_x(x,y)-A_x(x+1,y)\right] + \nonumber 
   \\
   &&-2 m_x(x,y-1)+{1\over \pi}\left[A_x(x,y)-A_x(x,y-1)\right] +\nonumber 
   \\
   &&-2 m_x(x,y+1)+{1\over \pi}\left[A_x(x,y)-A_x(x,y+1)\right]. 
			\label{s2}
\end{eqnarray}

\begin{eqnarray}
    {\partial U\over \partial m_y(x,y)}&=& -2 m_y(x,y-1)+{1\over 
    \pi}\left[A_y(x,y)-A_y(x,y-1)\right] +\nonumber \\ &&-2 
    m_y(x,y+1)+{1\over \pi}\left[A_y(x,y)-A_y(x,y+1)\right] +\nonumber 
    \\
    &&-2 m_y(x+1,y)+{1\over \pi}\left[A_y(x,y)-A_y(x+1,y)\right] +\nonumber 
    \\
    &&-2 m_y(x-1,y)+{1\over \pi}\left[A_y(x,y)-A_y(x-1,y)\right]. 
		    \label{s3}
\end{eqnarray}

\begin{ack}
	The authors thank Dr. G. Gonnella for useful discussions on 
	Mean-Field theory.
\end{ack}

\clearpage

\listoffigures 


\clearpage

\begin{thebibliography}{99}
\bibitem{Oppenheim-75}
    A. V. Oppenheim, R. W. Schafer, \emph{Digital Signal processing},
    Prentice-Hall, Englewood Cliffs, N.J., 1975.
\bibitem{Hjalmarson-85}
    H. P. Hjalmarson, L. A. Romero, D. C. Ghiglia, E. D. Jones, C. B. Norris,
    \emph{Phys. Rev. B} 32, 4300 (1985).
\bibitem{Nakadate-85}
    S. Nakadate, H. Saito, \emph{Applied Optics} 24, 2172 (1985).
\bibitem{Fried-77}
    D. L. Fried, \emph{J. Opt. Soc. Am.} 67, 370 (1977).
\bibitem{Ching-92}
    N. H. Ching, D Rosenfeld, M. Braun, \emph{IEEE Trans. Im. Proc.} 1, 355
    (1992).
\bibitem{Zebker-86}
    H. A. Zebker, R. M. Goldstein, \emph{J. Geophis. Res.} 91, (B5), 4993
    (1986).
\bibitem{Hadamard-02}
    J. Hadamard, \qt{Sur les probl\'emes aux d\'eriv\'ees partielles et leur
    signification physique}, \emph{Princeton University Bulletin} 13 (1902).
\bibitem{Geman-84}
    S. Geman, D. Geman, \emph{IEEE Trans. Pattern Anal. Mach. Intell.} 6, 721
    (1984).
\bibitem{Gidas-89}
    B. Gidas, \emph{IEEE Trans. Pattern Anal. Mach. Intell.} 11, 164 (1989).
\bibitem{Zhang-92}
    J. Zhang, \emph{IEEE Trans. Signal Process.} 40, 2570 (1992).
\bibitem{Tanaka-95}
    K. Tanaka, T. Morita, \emph{Phys. Lett. A} 203, 122 (1995).
\bibitem{Morita-96}
    T. Morita, K. Tanaka, \emph{Physica A} 223, 244 (1996).
\bibitem{Ray-97}
    J. Ray, R. W. Harris, \emph{Phys. Rev. E} 55, 5270 (1997).
\bibitem{Guerriero-98}
    L. Guerriero, G. Nico, G. Pasquariello, S. Stramaglia,
    \emph{Applied Optics} 37, 3053 (1998).
\bibitem{Stramaglia-99}
    S. Stramaglia, L. Guerriero, G. Pasquariello, N. Veneziani,
    \emph{Applied Optics} 38, 1377 (1999).
\bibitem{Ghiglia-98} 
    D. C. Ghiglia, M. D. Pritt,
    \emph{Two-Dimensional Phase Unwrapping. Theory, Algorithms, and Software},
    John Wiley \& Sons, New York, (1998). 
\bibitem{Ghiglia-94}
    D. C. Ghiglia, J. A. Romero, 
    \emph{J. Opt. Soc. Am. A} 11, 107--117, 1994.
\bibitem{Parisi-88}
    G. Parisi, \emph{Statistical Field Theory}, Addison-Wesley, Reading
    MA, (1988).
\bibitem{Yuille-94}
    For a review on Mean-Field Annealing methods, see e.g.\ A. L. 
    Yuille, J. J. Kosowsky, \emph{Neural Computation} 6, 341 (1994).
\end{thebibliography}
\end{document}